\documentclass[aps,amssymb,amsmath,pra,reprint,noshowpacs,superscriptaddress]{revtex4-1}
\usepackage{times}
\usepackage{amssymb,amsmath,graphicx}
\usepackage[usenames]{color}
\usepackage{bbold}

\usepackage[T1]{fontenc}
\usepackage[utf8]{inputenc}
\usepackage[english]{babel}

\usepackage{textcomp}

\newcommand{\diff}{\text{d}}
\newcommand{\vect}[1]{\boldsymbol{#1}}

\newcommand{\abs}[1]{\left| #1 \right|} 
\newcommand{\imu}{\text{i}}

\renewcommand{\d}[2]{\frac{\diff #1}{\diff #2}} 
\newcommand{\dd}[2]{\frac{\diff^2 #1}{\diff #2^2}} 

\newcommand{\commentOut}[1]{}
\usepackage{scalefnt}   

\newcommand{\affil}{Photonics Laboratory, ETH Zürich, 8093 Zürich, Switzerland}
\newcommand{\affilIhn}{Solid State Physics Laboratory, ETH Zürich, 8093 Zürich, Switzerland}

\begin{document}
\scalefont{1.05}
\title{A levitated nanoparticle as a classical two-level atom}
\author{Martin Frimmer}
\affiliation{\affil}
\homepage{http://www.photonics.ethz.ch}
\author{Jan Gieseler}
\affiliation{\affil}
\author{Thomas Ihn}
\affiliation{\affilIhn}
\author{Lukas Novotny}
\affiliation{\affil}

\begin{abstract}
The center-of-mass motion of a single optically levitated nanoparticle resembles three uncoupled harmonic oscillators. We show how a suitable modulation of the optical trapping potential can give rise to a coupling between two of these oscillators, such that their dynamics are governed by a classical equation of motion that resembles the Schrödinger equation for a two-level system. Based on experimental data, we illustrate the dynamics of this parametrically coupled system both in the frequency and in the time domain. We discuss the limitations and differences of the mechanical analogue in comparison to a true quantum mechanical system.
\end{abstract}
\date\today

\maketitle

\section{Introduction}
One particularly interesting laboratory system for engineers and physicists alike are optically levitated nanoparticles in vacuum~\cite{Ashkin1977,Li2011,Gieseler2012}. The fact that such a levitated particle at suitably low pressures is a harmonic oscillator with no mechanical interaction with its environment promises exceptional control over the system's dynamics~\cite{Millen2015,Kiesel2013}. In fact, at such low pressures, the only interaction expected to remain between the particle and the environment is mediated by the electromagnetic field~\cite{Chang2010,Romero-Isart2011}. Indeed, it has recently been shown that at sufficiently low pressures, the dominating interaction of the particle with its environment is determined by the photon bath of the trapping laser~\cite{Jain2016}. Therefore, with the understanding of quantum electrodynamics gained by quantum optics and atomic physics over the last decades, optically levitated nanoparticles are relatively massive mechanical systems whose decoherence might be controlled to an unprecedented level.
While the dynamics of optically levitated particles have been controlled to a remarkable degree, it is surprising that little attention has been paid to the fact that a single optically levitated nanoparticle is an embodiment of three harmonic oscillators, one for each degree of freedom of the particle's center-of-mass motion, offering the opportunity for introducing a coupling between these modes~\cite{Blaum2008,Cornell1990}.
For clamped-beam micromechanical systems, the coupling between different oscillation modes has been explored in great detail~\cite{Venstra2011,Faust2013,Matheny2013,Gavartin2013,Okamoto2013,Mahboob2014}.
In contrast, for optically levitated particles, only recently the first steps have been taken to couple different degrees of freedom of the center-of-mass motion and harness the machinery of coherent control established on quantum-mechanical systems~\cite{Frimmer2016}.

In this paper, we demonstrate explicitly how a simple modulation of the optical trapping potential couples two degrees of freedom of a levitated nanoparticle, whose dynamics is then governed by an equation of motion resembling the Schrödinger equation of a quantum-mechanical two-level system. We experimentally investigate the scaling of the coupling frequency with coupling strength and detuning. In our discussion, we focus on the complementary observations of the dynamics of the particle in the frequency domain and in the time domain. Finally, we discuss the limitations of our model and those of the analogy between a classical two-mode system and a quantum-mechanical two-level atom.

\section{Experimental system}
\begin{figure}[btp]
\centering
{\includegraphics[width=\linewidth]{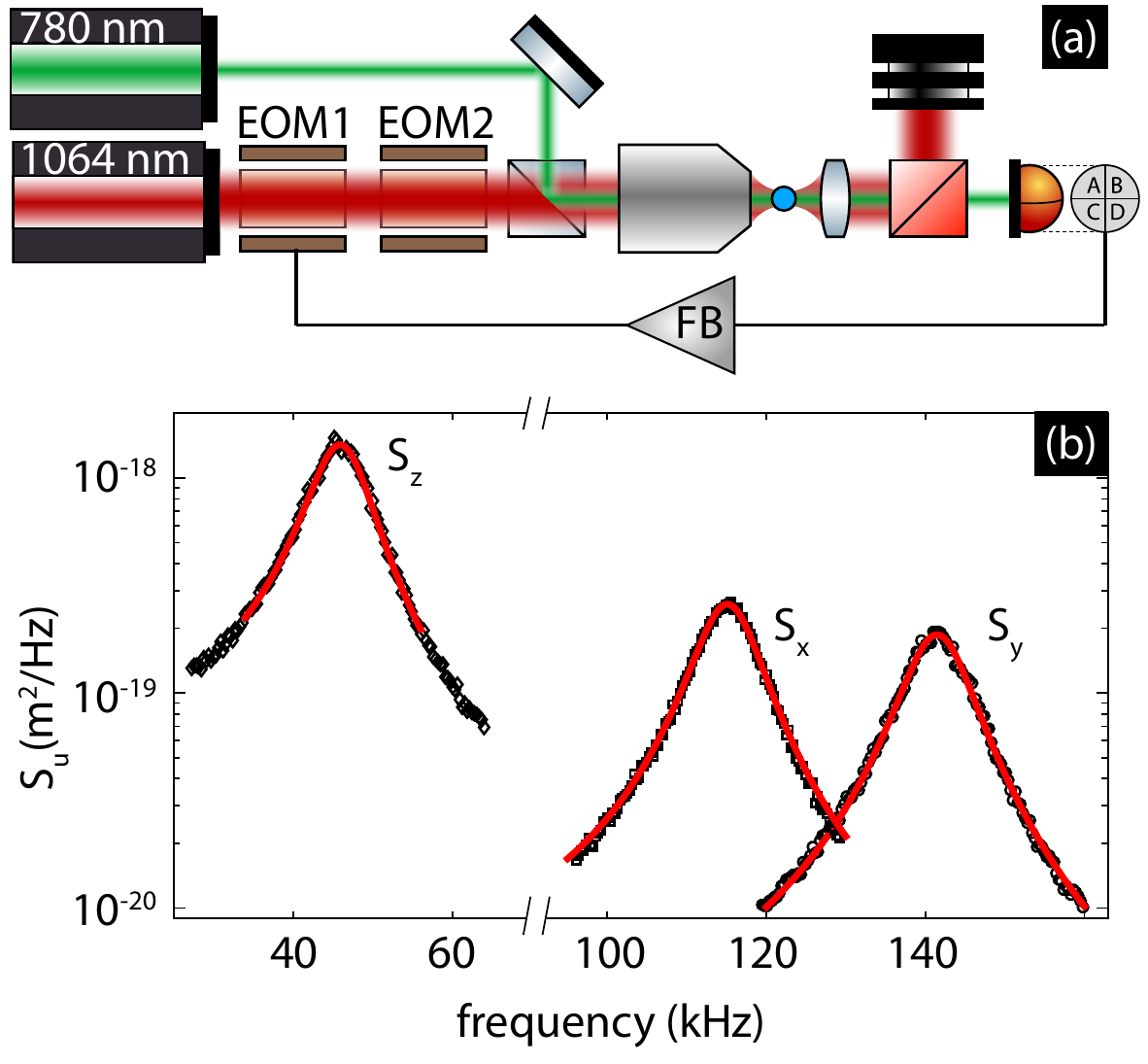}}
\caption{(a) Experimental setup. The trapping laser (1064\,nm) passes two electro-optical modulators for intensity and polarization control, before it is focused by a microscope objective. A measurement beam (780\,nm) is coaligned with the trapping laser to measure the particle's position in a quadrant detection scheme. A feedback system modulates the intensity of the trapping laser. (b) Power spectral densities $S_x$, $S_y$, and $S_z$ of the particle motion in $x$, $y$, and $z$ direction, respectively, recorded at 10\,mbar.}
\label{fig:setup_powerSpectra}
\end{figure}
At the heart of our experimental setup is a single-beam optical dipole trap, illustrated in Fig.~\ref{fig:setup_powerSpectra}(a). The trapping laser (wavelength $1064$\,nm, power $\approx50$\,mW) is focused by a microscope objective (100$\times$, NA0.9) to a diffraction limited spot forming the trap. We use two electro-optic modulators (EOM) to condition the trapping laser. The first modulator (EOM1) together with a polarizing beam splitter [not shown in Fig.~\ref{fig:setup_powerSpectra}(a)] serves to modulate the laser intensity. The trapping laser then enters a second modulator (EOM2). The voltage applied to this modulator rotates the polarization direction of the laser beam.
We trap silica particles with a nominal diameter of 136\,nm in the laser focus. Particles much smaller than the wavelength of the trapping laser are well described in the dipolar approximation, where a scatterer in a focused laser field is subjected to two forces~\cite{Novotny2012}. First, the gradient force pulls a dielectric scatterer along the gradient of the intensity to the focal center. Second, the scattering force pushes the particle along the propagation direction of the laser beam. Since the gradient force scales with the particle volume, while the scattering force goes with the square of the volume, sufficiently small particles can be stably trapped in a conservative potential dominated by the gradient force. With the origin chosen to be the center of the focus, the trapping potential is harmonic to lowest order in the particle position, such that the equations of motion of a trapped particle are those of a three-dimensional harmonic oscillator.

\begin{figure}[tpb]
\centering
\includegraphics[width=\linewidth]{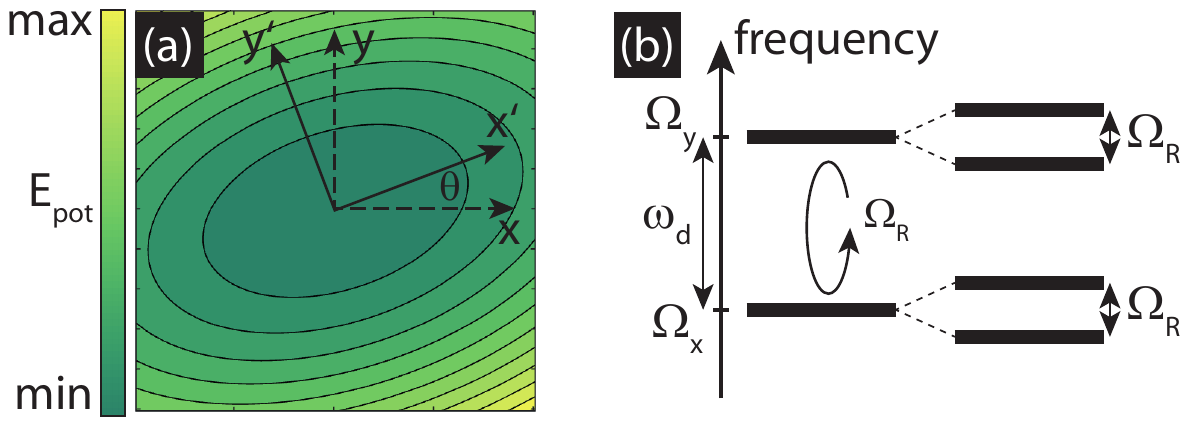}
\caption{(a)~Schematic illustration of the parabolic trapping potential in false color. The coordinate system $(x',y')$ is aligned with the main axes of the potential, while the coordinate system $(x,y)$ is not.
(b)~Level scheme of the mechanical atom. The two bare eigenmodes of the trapped particle have eigenfrequencies $\Omega_x$ and $\Omega_y$, respectively. In the presence of a driving field, corresponding to a periodic rotation of the potential by a small angle in our case, the modes are coupled at a frequency $\Omega_R$. This coupling gives rise to two doublets of dressed states, separated by the Rabi frequency $\Omega_R$.}
\label{fig:rotatingPotential}
\end{figure}

In order to detect the particle position, we employ a second laser beam (wavelength $780$\,nm, power $\approx3$\,mW), sufficiently weak to not disturb the potential created by the trapping laser. The light scattered by the trapped particle is collimated by a collection lens, spectrally filtered to remove the trapping laser, and guided to a split-detection setup to infer the position of the particle as a function of time.
A convenient way to characterize the trapping potential is to observe the particle motion under the stochastic driving force arising from the interaction with the surrounding gas molecules. In Fig.~\ref{fig:setup_powerSpectra}(b), we plot the power spectra $S_x(f)$, $S_y(f)$ and $S_z(f)$ of the particle position along $x$, $y$, and $z$, respectively, recorded at a pressure of $10$~mbar. We denote with $z$ the position along the optical axis, and with $x$ and $y$ the position in the focal plane relative to the focal point. Each spectrum is a Lorentzian function (see fits to data), each defined by three parameters. The center frequency $\Omega_u$ is given by the trap stiffness in the respective direction $u\in\left\{x,y,z\right\}$. This stiffness scales with the trapping laser power. The width of the Lorentzian is set by the damping rate $\gamma$, which scales linearly with gas pressure. The area under the power spectrum by definition equals the variance of the position $\langle u^2\rangle= \int_0^\infty\diff f\,S_u(f)$, which has to satisfy $\langle u^2\rangle=k_BT/(m\Omega_u^2)$ according to the equipartition theorem, where $k_B$ is the Boltzmann constant, $T$ is the temperature of the bath (which is room temperature in our case), and $m$ is the mass of the particle (which is nominally $2.9\times10^{-19}$\,kg). In fact, it is the equipartition theorem which allows us to convert the voltage output by our detectors to a position in meters. The oscillation frequencies found experimentally in Fig.~\ref{fig:setup_powerSpectra}(b) are $\Omega_z=2\pi\times46$\,kHz, $\Omega_x=2\pi\times115$\,kHz, and $\Omega_y=2\pi\times141$\,kHz. The oscillation frequency along the optical axis is significantly lower than those in the transverse plane, since the focal depth of a standard high-NA objective exceeds the transverse confinement of the focal field. Importantly, the rotational symmetry of the trapping optics is broken by the linear polarization of the laser beam, which leads to an intensity distribution in the focus, illustrated in Fig.~\ref{fig:rotatingPotential}(a), which is elongated along the direction of polarization, lifting the degeneracy of the eigenfrequencies of the motion in the focal plane.

The non-degeneracy of all oscillation modes allows us to control the three degrees of freedom of the particle's center-of-mass motion simultaneously~\cite{Gieseler2012}. We use analog electronics to generate a signal which is proportional to $u\dot u$, which we feed back to EOM1 to modulate the intensity of the trapping laser, which in turn modulates the trap stiffness at a frequency $2\Omega_u$~\cite{Gieseler2012}. At sufficiently low pressures, the frequency spacing between the oscillation modes largely exceeds their spectral width, such that each mode only responds to the feedback signal generated from its own position time trace. Accordingly, under feedback, we are dealing with three uncoupled, parametrically driven harmonic oscillators. By adjusting the phase of the feedback signal, we can choose to parametrically heat the respective center-of-mass mode, increasing its amplitude, or to parametrically cool the mode, reducing its amplitude below the thermal population~\cite{Gieseler2014}. In good approximation, under feedback cooling, the particle appears to be coupled to an effective bath at temperature $T_\mathrm{eff}$, which is below room temperature.
Throughout this paper, we feedback-cool the $z$-mode of the particle to a center-of-mass temperature around 1\,K, effectively restricting the particle motion to the focal plane. Note that, in principle, one can avoid the separate measurement laser and use the scattering of the trapping laser to measure the particle position. However, this approach requires more sophisticated filtering of the detector signal to avoid the feedback modulation of the trapping laser reentering the feedback loop~\cite{Jain2016}.

\section{Coupling the particle's transverse modes}
Let us now investigate the dynamics of the particle as the polarization of the trapping laser is modulated. Considering the $z$-mode to be frozen out by feedback-cooling, the trapping potential effectively reads
\begin{equation}
\label{eq:pot}
V=\frac{k-\Delta k}{2}x'^2+\frac{k+\Delta k}{2}y'^2,
\end{equation}
with $k-\Delta k=m\Omega_x^2$ and $k+\Delta k=m\Omega_y^2$. Here, we have denoted the coordinates of the particle in the focal plane with $x'$ and $y'$ for later convenience and assumed that these coordinate axes are aligned with the main axes of the parabolic potential, as illustrated in Fig.~\ref{fig:rotatingPotential}(a).
In a rotated coordinate system $(x,y)$ with
\begin{equation}
\begin{aligned}
\label{eq:coordTrafo}
x'&=\phantom{-} \cos(\theta)x+\sin(\theta)y,\\
y'&=-\sin(\theta)x+\cos(\theta)y,
\end{aligned}
\end{equation}
the potential takes the form
\begin{equation}
\label{eq:potTrafo2}
\begin{aligned}
V&=\frac{k-\Delta k\cos(2\theta)}{2}x^2+\frac{k+\Delta k\cos(2\theta)}{2}y^2\\
&~~~-xy\Delta k\sin(2\theta).
\end{aligned}
\end{equation}
Assume now that we periodically rotate the potential around the $z$-axis by a small angle $\delta$ at the driving frequency $\omega$ with a phase $\varphi$, such that we have
\begin{equation}
\label{eq:variationTheta}
\theta=\delta\cos(\omega t+\varphi).
\end{equation}
To linear order in $\theta$ we find the potential
\begin{equation}
\label{eq:potFirstOrder}
V_\text{lin}=\frac{k-\Delta k}{2}x^2+\frac{k+\Delta k}{2}y^2-2\delta\cos(\omega t+\varphi)\Delta k ~xy.
\end{equation}
With the carrier frequency $\Omega_0=k/m$, the frequency splitting $\Omega_d^2=\Delta k/m$, and the coupling frequency $\Omega_\delta^2=\Delta k\delta/m=\Omega_d^2\delta$ we find the following equations of motion~\cite{Frimmer2014}
\begin{equation}
\begin{aligned}
\label{eq:coupledoscMatrixFormEigen}
  \begin{bmatrix} \dd{}{t} + \gamma \d{}{t} + \Omega_0^2-\Omega_d^2 & -2\Omega_\delta^2\cos(\omega t+\varphi)\\ -2\Omega_\delta^2\cos(\omega t+\varphi) & \dd{}{t} + \gamma \d{}{t} + \Omega_0^2+\Omega_d^2\end{bmatrix}\begin{bmatrix} x \\ y \end{bmatrix}=\vect{F},
\end{aligned}
\end{equation}
where $\vect{F}$ is a force driving the system.
Here, we have introduced the damping rate $\gamma$ for both oscillators. We introduce the complex amplitudes $a(t)$ and $b(t)$ for the oscillation along $x$ and $y$, respectively, by writing
\begin{equation}
\label{eq:SVEAansatz}
\begin{aligned}
x & \,=\,\text{Re}\left\{a(t)\exp{[\imu\Omega_0t]}\right\},\\
y & \,=\,\text{Re}\left\{b(t)\exp{[\imu\Omega_0t]}\right\}.
\end{aligned}
\end{equation}
In the slowly varying envelope approximation (neglecting second derivatives of the amplitudes with respect to time, and considering strongly underdamped oscillators, such that we have $2\imu\Omega_0+\gamma\approx 2\imu\Omega_0$), we obtain for the equations of motion for the oscillation amplitudes in the absence of a driving force
\begin{equation}\label{eq:SVEAsystem}
\imu\begin{bmatrix} \dot{a} \\ \dot{b}\end{bmatrix} \;=\; \frac{1}{2}\begin{bmatrix} \omega_d-\imu\gamma & 2\omega_\delta\cos(\omega t+\varphi)\\  2\omega_\delta\cos(\omega t+\varphi) & -\omega_d-\imu\gamma \end{bmatrix} \begin{bmatrix} a\\b\end{bmatrix},
\end{equation}
where we have introduced the rescaled driving frequency $\omega_\delta=\Omega_\delta^2/\Omega_0$, as well as the rescaled frequency splitting $\omega_d=\Omega_d^2/\Omega_0$.
In the limit of vanishing damping $\gamma$, the system of equations for the complex mode amplitudes $a$ and $b$ exactly resembles the Schrödinger equation for a two-level system, whose levels with complex amplitudes $a$ and $b$ are split in energy by $\hbar\omega_d$ and are coupled at the rate $\omega_\delta$ by a driving field oscillating at frequency $\omega$~\cite{Allen1987}.
To solve this Rabi problem, we change to a frame rotating at the driving frequency with the transformation
\begin{equation}
\begin{aligned}
\label{eq:RotFrameAnsatz}
a&\;=\;\bar{a}(t)\exp{[-\imu\frac{\omega}{2}t]},\\
b&\;=\;\bar{b}(t)\exp{[+\imu\frac{\omega}{2}t]},
\end{aligned}
\end{equation}
and apply the rotating-wave approximation (neglecting counter-rotating terms) to obtain
\begin{equation}\label{eq:ihdt_=H}
\imu\d{}{t}\begin{bmatrix} {\bar{a}} \\ {\bar{b}}\end{bmatrix} \;=\; H \begin{bmatrix} \bar{a}\\\bar{b}\end{bmatrix},
\end{equation}
with the coupling matrix
\begin{equation}\label{eq:classicalHamiltonian}
  H = \frac{\Delta-\imu\gamma}{2}\sigma_z + \frac{\omega_x}{2}\sigma_x+\frac{\omega_y}{2}\sigma_y,
\end{equation}
and the detuning $\Delta=\omega_d-\omega$ of the driving field relative to the level splitting, the Pauli spin matrices $\sigma_i$, and the coupling rates $\omega_x=\omega_d\delta\cos\varphi$ and $\omega_y=-\omega_d\delta\sin\varphi$.

Let us summarize our findings thus far. We have considered the center-of-mass motion of an optically levitated nanoparticle in the focal plane of an optical dipole trap. A periodic rotation of the trapping potential around the optical axis by a small angle leads to a parametric coupling between the two in-plane modes. In the limit of vanishing damping $\gamma$, the time-dependent energy in the modes, which is proportional to $\abs{a}^2$ and $\abs{b}^2$, respectively, follows the same dynamics as the populations of a two-level atom driven by a classical light field. Figure~\ref{fig:rotatingPotential}(b) illustrates the level scheme of our mechanical atom. The two bare eigenmodes of the particle have frequencies $\Omega_x$ and $\Omega_y$, respectively. The level splitting of the mechanical atom is given by $\omega_d$, which equals $\Omega_y-\Omega_x$ in the limit of the carrier frequency $\Omega_0$ largely exceeding the splitting $\omega_d$. It is well-known that the populations of a two-level system under a near-resonant drive undergo oscillations at the generalized Rabi frequency
\begin{equation}\label{eq:RabiFreq}
\Omega_R=\sqrt{\omega_\delta^2+\Delta^2}.
\end{equation}

\section{Experimental results}
\begin{figure}[tb]
\centering
{\includegraphics[width=\linewidth]{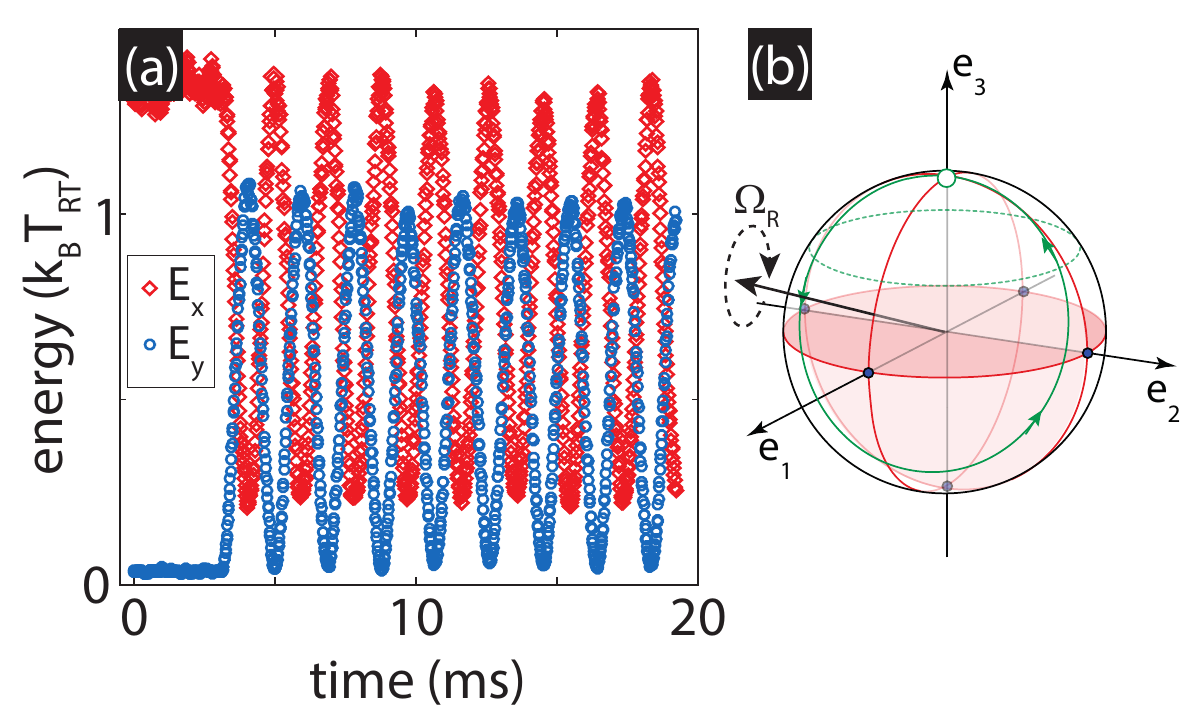}}
\caption{(a)~Classical Rabi oscillations in the energy of the $x$-mode (red diamonds) and the $y$-mode (blue circles). The system is initialized with the $x$-mode heated to $1.4\,k_BT$ and the $y$-mode cooled to $0.04\,k_BT$. The feedback is switched off during the entire time shown. At $t=3\,$ms, the coupling is turned on and energy is periodically exchanged between the modes.
(b)~Bloch sphere representation of the measurement in (a). The system is initialized close to the north pole of the Bloch sphere. When the coupling is turned on, the Bloch vector rotates around the rotation vector, whose component in the equatorial plane is given by the driving strength $\omega_\delta$, and whose $z$-component is set by the detuning $\Delta$.}
\label{fig:RabiOscillationsBlochSphere}
\end{figure}

A measurement of such a classical Rabi experiment is shown in Fig.~\ref{fig:RabiOscillationsBlochSphere}(a). We initialize the system by feedback-cooling the $y$-mode to a center-of-mass temperature of $0.04k_BT$. Using parametric driving we have moderately heated the $x$-mode to $1.4k_BT$. At time $t=0$ any modulation of the trapping potential is switched off and the modes are freely evolving. The experiment is conducted at a pressure of $5\times10^{-6}$~mbar, where the inverse damping rate of the particle is several orders of magnitude longer than the duration of the experiment. At time $t=3$\,ms we start to modulate the polarization direction of the trapping laser at a frequency $\omega=2\pi\times28.7$\,kHz, which is very close to the frequency splitting of the $x$- and $y$-mode. We observe oscillations of the energy between the two oscillation modes of the particle in the focal plane at a frequency $\Omega_R=2\pi\times540$\,Hz.

We can conveniently illustrate the behavior of our system on the Bloch sphere, shown in Fig.~\ref{fig:RabiOscillationsBlochSphere}(b). When all energy in the system resides in the $x$-mode ($y$-mode), the Bloch vector describing the system points to the north pole (south pole) of the sphere. Points on the equator denote states with equal amplitude in both modes, where the relative phase between the modes determines the location along the equator. Accordingly, the measurement shown in Fig.~\ref{fig:RabiOscillationsBlochSphere}(a) starts close to the north pole. Turning on the driving generates a rotation vector, around which the Bloch vector rotates at a frequency $\Omega_R$.
The component of the rotation vector in the equatorial plane is given by the coupling frequency $\omega_\delta$, while the out of plane component is determined by the detuning $\Delta$, which therefore governs the contrast of the energy transfer in Fig.~\ref{fig:RabiOscillationsBlochSphere}(a).

\begin{figure}[t]
\centering
{\includegraphics[width=\linewidth]{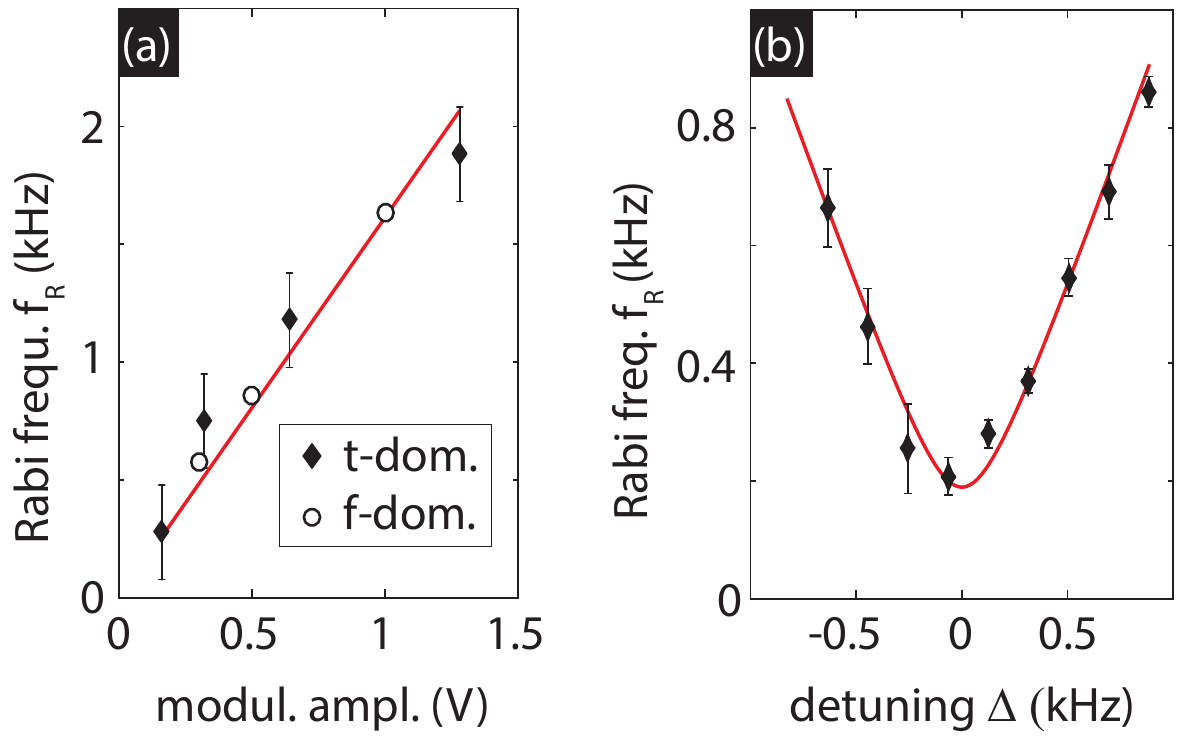}}
\caption{(a)~Scaling of the Rabi frequency with driving strength, expressed as the modulation amplitude applied to EOM2. The filled diamonds represent data extracted from Rabi oscillations observed in the time domain, similar to the measurement shown in Fig.~\ref{fig:RabiOscillationsBlochSphere}(a). The open circles are extracted from an eigenmode analysis in the presence of thermal fluctuations, as shown in Fig.~\ref{fig:anticrossing}(b). The red line is a linear fit to the data. (b)~Measured scaling of the Rabi frequency as a function of detuning $\Delta$ of the driving frequency relative to the level splitting $\omega_d$. The red line illustrates the scaling according to Eq.~\eqref{eq:RabiFreq}.}
\label{fig:scalingRabi}
\end{figure}

We have experimentally investigated the scaling of the Rabi frequency as a function of driving strength, which is set by the angle $\delta$ by which the potential is rotated, and, in our case, scales linearly with the voltage applied to EOM2. In the absence of detuning, the Rabi frequency scales linearly with driving strength, according to Eq.~\eqref{eq:RabiFreq}. In Fig.~\ref{fig:scalingRabi}(a), we plot as solid diamonds the Rabi frequency extracted from the oscillations in the population $\abs{a(t)}^2$ and $\abs{b(t)}^2$ for four different driving strengths, given as the amplitude of the sinusoidal voltage applied to EOM2. The experimental results are in good agreement with the expected linear scaling. In Fig.~\ref{fig:scalingRabi}(b) we experimentally investigate the scaling of the Rabi frequency with detuning $\Delta$. The data corresponds well with the theoretical expectation according to Eq.~\eqref{eq:RabiFreq}.

\begin{figure}[t]
\centering
\includegraphics[width=\linewidth]{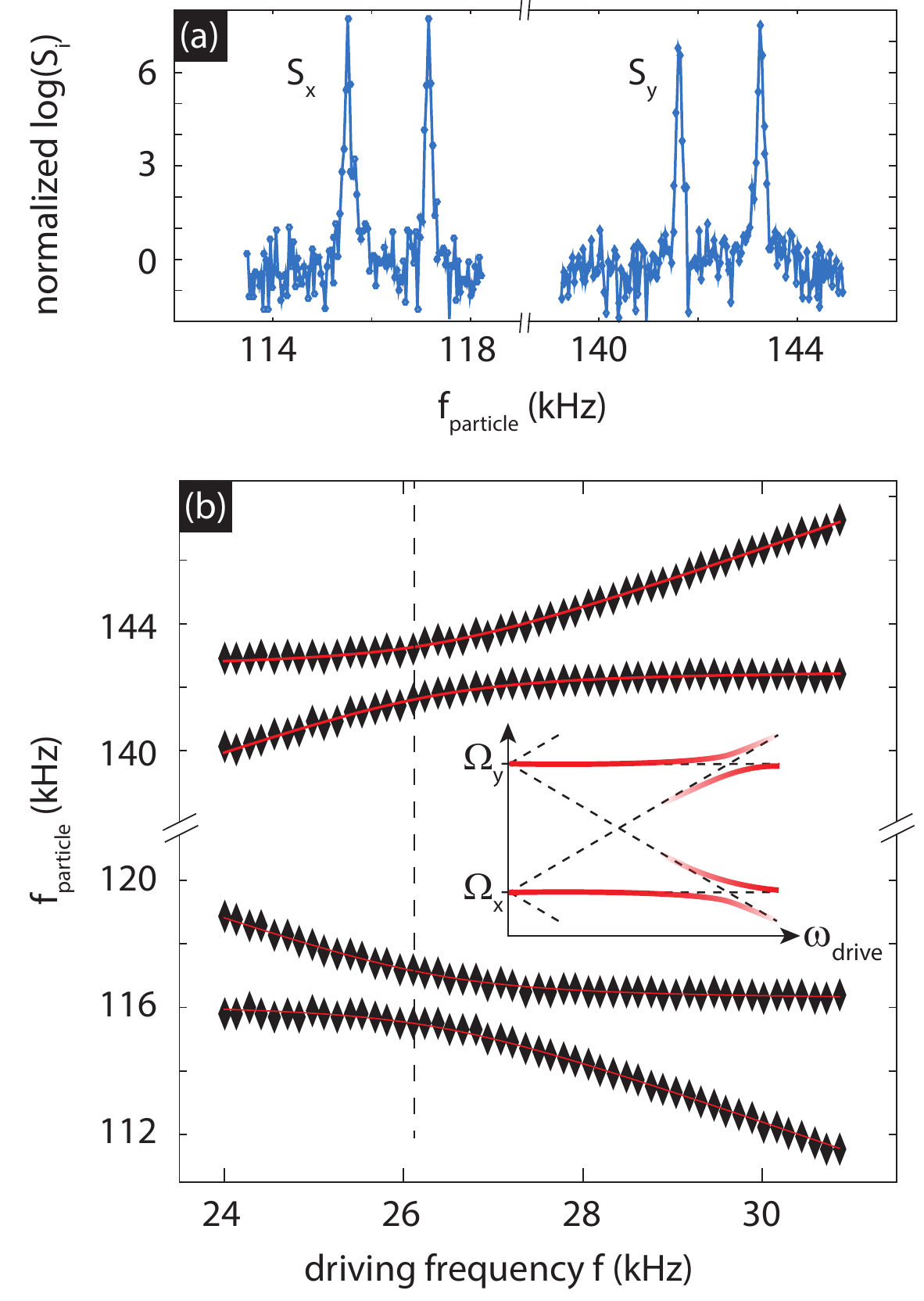}
\caption{(a) Power spectral densities of detector signals $S_x$ and $S_y$ under feedback-cooling and thermal driving with the coupling between the modes turned on. The signal is normalized to the detector noise floor. Each oscillation mode is split into two dressed modes.
(b)~Frequencies of the dressed modes as a function of driving frequency. The hybrid modes show an anticrossing. Inset: The hybrid modes can be understood as arising due to the coupling of one sideband of each bare mode, generated by the modulation of the trapping potential, hybridizing with the other bare mode.}
\label{fig:anticrossing}
\end{figure}

So far, we have discussed the signatures of the coupling between the two oscillation modes of the trapped particle in the focal plane in the time domain by observing the temporal evolution of the mode temperatures, which are proportional to the populations $\abs{a(t)}^2$ and $\abs{b(t)}^2$. It is instructive to take a complementary view at the dynamics of the parametrically coupled system in the frequency domain. Let us recall that, as shown in Fig.~\ref{fig:setup_powerSpectra}(b), the power spectral density of the particle's motion, driven by the white noise of the thermal fluctuations of the bath, renders a visualization of the mode distribution in frequency space. When the driving field coupling the two oscillation modes of the particle in the focal plane is switched off, we found one Lorentzian mode for each degree of freedom of the particle [Fig.~\ref{fig:setup_powerSpectra}(b)]. Interestingly, when the driving at a frequency close to resonance ($\Delta\approx0$) is turned on, each mode in the focal plane splits into a doublet of dressed states, as shown in Fig.~\ref{fig:anticrossing}(a) and schematically illustrated in Fig.~\ref{fig:rotatingPotential}(b). The data was acquired at a pressure of $5\times10^{-6}$~mbar. The linewidth of the hybrid modes in Fig.~\ref{fig:anticrossing}(a) is Fourier limited by the acquisition time. Clearly, our system is located well in the strong coupling regime, where the dressed-mode splitting, which is the classical analogy of the Autler-Townes splitting~\cite{Fox2006}, exceeds the linewidth. The Rabi oscillations observed in the populations $\abs{a(t)}^2$ and $\abs{b(t)}^2$ in the time domain in Fig.~\ref{fig:RabiOscillationsBlochSphere}(a) are simply the beating between these hybrid modes and, accordingly, the dressed-mode splitting equals the Rabi frequency.
In Fig.~\ref{fig:anticrossing}(b), we plot the measured frequencies of the dressed modes as a function of the frequency of the driving field. The spectrum shown in Fig.~\ref{fig:anticrossing}(a) corresponds to the dashed line in Fig.~\ref{fig:anticrossing}(b). We observe a characteristic anticrossing of the hybrid modes as the driving frequency is swept across resonance $\omega=\Omega_y-\Omega_x$. The inset of Fig.\ref{fig:anticrossing}(b) shows a schematic illustration of the mode structure of the parametrically coupled system. The modulation of the coupling at frequency $\omega$ generates sidebands on the bare eigenmodes. It is the upper sideband $\Omega_x+\omega$ of the lower bare mode $\Omega_x$ which hybridizes with the bare mode at $\Omega_y$ to form the upper doublet of dressed modes, while the lower sideband $\Omega_y-\omega$ of the upper mode hybridizes with the bare mode $\Omega_x$ to generate the lower doublet. The non-resonant sidebands $\Omega_x-\omega$ and $\Omega_y+\omega$ are neglected in the analysis when applying the rotating-wave approximation.
Our coupled-mode theory Eq.~\eqref{eq:ihdt_=H} yields the dressed-mode frequencies
\begin{equation}\label{}
  \begin{aligned}
    \Omega_x^\pm = \Omega_0+\frac{\omega}{2}\pm\frac{\Omega_R}{2},\\
    \Omega_y^\pm = \Omega_0-\frac{\omega}{2}\pm\frac{\Omega_R}{2},
  \end{aligned}
\end{equation}
which fits our data well [red lines in Fig.~\ref{fig:anticrossing}(b)]. From the fit we extract the resonant Rabi frequency $\Omega_R(\Delta=0)$ for three different driving strengths and add the result to the plot in Fig.~\ref{fig:scalingRabi}(a) as the open symbols. The data extracted from the eigenmode analysis in the frequency domain corresponds well to that obtained from time domain measurements of the Rabi oscillations (black diamonds). We note that the dependence of the Rabi frequency on the detuning $\Delta$ investigated in Fig.~\ref{fig:scalingRabi}(b) is reproduced by the dependence of the frequency splitting of the dressed modes on the driving frequency found in Fig.~\ref{fig:anticrossing}(b).

\section{Discussion}
Let us summarize the similarities of our classical two-mode system to a quantum-mechanical two-level system. We have found that the complex slowly varying amplitudes of the parametrically coupled classical harmonic oscillators follow an equation of motion that has the form of the Schrödinger equation for the complex amplitudes of a quantum-mechanical two-level system. Accordingly, we can apply the machinery of coherent control, regularly used to control qubits, to classical harmonic oscillator systems~\cite{Okamoto2013,Faust2013,Frimmer2016}.
For completeness, we point out some limitations of the mechanical atom. There are two limitations to the driving strength and thereby to the reachable Rabi frequency. The first limitation is set by the carrier frequency $\Omega_0$ of the oscillators and concerns the step from a Newtonian equation of motion, which is of second order in time, to a coupled mode equation, which is of first order in time. The transition from Eq.~\eqref{eq:coupledoscMatrixFormEigen} to Eq.~\eqref{eq:SVEAsystem} required the slowly varying envelope approximation, which is clearly violated when the temporal variation of the amplitudes $a(t)$ and $b(t)$, given by the Rabi frequency $\Omega_R$, becomes comparable to the carrier frequency $\Omega_0$. The second limitation regards our specific experimental embodiment of the mechanical atom and concerns the coupling strength $\Omega_R$ achievable relative to the level splitting $\Omega_y-\Omega_x$. The coupling frequency $\omega_\delta$ entering Eq.~\eqref{eq:SVEAsystem} is proportional to the level splitting $\omega_d$ and the rotation angle of the potential $\delta$. In order for our linear approximation of the potential in Eq.~\eqref{eq:potFirstOrder} to hold, the rotation angle has to fulfill the condition $\delta\ll1$, which is turn means that in this limit the Rabi frequency can never become comparable to the level splitting and our system cannot enter the regime of ultra-strong coupling where the rotating-wave approximation breaks down.
Even more important than these practical limitations are the differences between our classical analogue and the quantum-mechanical two-level system. A clear signature of the classical nature of our system is the absence of Planck's constant. Of course we could multiply both sides of Eq.~\eqref{eq:SVEAsystem} with $\hbar$ and express all frequencies in the coupling matrix as energies. However, this operation cannot mask the fact that our classical theory neither implies any correspondence between energy and frequency, nor does it require any discrete gridding of phase space.
More importantly, our mechanical atom illustrates the well known fact that the miraculous nature of quantum mechanics is not captured by the Schrödinger equation. The quantization of states into discrete modes, the time evolution of these modes under an equation first order in time, interference between these modes, and uncertainty relations between Fourier-conjugate variables are features characteristic for any wave theory and not special to quantum mechanics. Instead, the essence of quantum mechanics comes with the Copenhagen interpretation, for example predicting the collapse of the wavefunction under a projective measurement. In our classical system, no such collapse exists. For example, in Fig.~\ref{fig:RabiOscillationsBlochSphere}(a), we continuously observe the population in the two oscillation modes as they undergo Rabi oscillations in a single experimental run. The analogous measurement on a quantum-mechanical two-level system would require multiple experiments with identically prepared systems to generate statistics, since the quantum-mechanical populations have to be interpreted as probabilities to find the system in the respective state. Our system nicely illustrates the fact that the collapse of the wavefunction under a projective measurement does not have to be interpreted as a nuisance. Instead, a projective measurement is a very convenient method to reliably initialize a quantum-mechanical system, a resource not available in the realm of classical physics. For example, initializing our classical two-level atom in some eigenstate, \emph{i.e.}, bringing its Bloch vector to one of the poles of the Bloch sphere, requires a sophisticated coherent control protocol instead of a single projective measurement~\cite{Frimmer2016}. Finally, we point out that our equation of motion for the mechanical atom Eq.~\eqref{eq:ihdt_=H} does not contain any effect resembling spontaneous emission. This fact does not harm the analogy to the quantum-mechanical two-level system, since also the Schrödinger equation does not include spontaneous transitions. In the presence of finite damping $\gamma$, however, the classical ``Hamiltonian'' coupling matrix Eq.~\eqref{eq:classicalHamiltonian} becomes non-hermitian and the total population of the mechanical atom is leaking out of the system.

\section{Conclusions}
In conclusion, we have presented a mechanical system of two parametrically coupled classical harmonic oscillators that follows an equation of motion which is formally equivalent to the Schrödinger equation describing the interaction of a quantum-mechanical two-level system interacting with a classical field. We have provided two complementary views on the dynamics of our system. The first view focused on the Rabi oscillations, transferring energy between the parametrically coupled oscillator modes in the time domain. The second view focused on the frequency spectrum of the coupled-mode system, where the driving field dresses the bare eigenmodes giving rise to a set of hybrid states, in analogy to the Autler-Townes splitting.
We have used our system to illustrate some properties it shares with a quantum-mechanical two-level atom, and to finally point out some unique features of quantum mechanics, which reach beyond the realm of our classical analogy.

This research was supported by ERC-QMES (Grant No. 338763) and the NCCR-QSIT program (Grant No. 51NF40-160591). J. G. has been partially supported
by the Postdoc-Program of the German Academic Exchange Service (DAAD) and H2020-MSCA-IF-2014 under REA grant Agreement No. 655369. We thank L. Rondin for valuable input and discussions.

\bibliography{Frimmer_Literature}

\end{document}